\documentclass[a4paper]{llncs}

\usepackage{amssymb}
\usepackage{graphicx}
\usepackage{booktabs}

\usepackage{url}
\urldef{\mailsa}\path|{herrmann, 0maass, federrath}@informatik.uni-hamburg.de|

\usepackage{array}

\usepackage[utf8]{inputenc}
\usepackage[T1]{fontenc}

\usepackage[protrusion=true,expansion=false]{microtype}

\usepackage{cleveref} 

\let\mySection\section\renewcommand{\section}{\suppressfloats[t]\mySection}

\usepackage{afterpage}

\begin{document}

\mainmatter
\title{Evaluating the Security of a DNS Query Obfuscation Scheme for Private Web Surfing\thanks{This is the author's version. The final publication is available at Springer via http://dx.doi.org/10.1007/978-3-642-55415-5\_17}}
\author{Dominik Herrmann \and Max Maaß \and Hannes Federrath}

\institute{University of Hamburg, Computer Science Department, Germany
}
\maketitle

\begin{abstract}

The Domain Name System (DNS) does not provide query privacy. Query obfuscation schemes have been proposed to overcome this limitation, but, so far, they have not been evaluated in a realistic setting. In this paper we evaluate the security of a random set range query scheme in a real-world web surfing scenario. We demonstrate that the scheme does not sufficiently obfuscate characteristic query patterns, which can be used by an adversary to determine the visited websites. We also illustrate how to thwart the attack and discuss practical challenges. Our results suggest that previously published evaluations of range queries may give a false sense of the attainable security, because they do not account for any interdependencies between queries.

\end{abstract}

\section{Introduction}
\label{sec:Introduction}

DNS is an integral part of the Internet infrastructure. Unfortunately, it does not offer privacy, i.\,e., the so-called resolvers (recursive nameservers) can see all queries sent to them in the clear. Resolvers can learn about users' habits and interests, which may infringe their privacy if the resolver is not run by a trusted party, but by a third party such as Google, whose resolver 8.8.8.8 serves more than 130 billion queries per day on average \cite{googledns}. The discussions about limiting tracking via cookies spurred by the ``Do not track'' initiative may result in DNS queries becoming the next target for tracking and profiling purposes \cite{Conrad12-dnssecurity}. According to \cite{HBF:2013} behavior-based tracking based on DNS queries may be feasible.

Integrating mechanisms for confidentiality into DNS is difficult because of the need for compatibility with existing infrastructure. Fundamental changes to the protocol are implemented very slowly, as previous attempts have shown: Although the initial DNSSEC security extensions have been proposed in 1999 \cite{rfc2535}, the majority of users still can not profit from their benefits today. Unfortunately, DNSSEC does not address privacy issues due to an explicit design decision \cite{rfc4033}.

Currently, there is no indication that facilities for privacy-preserving resolution will be integrated into the DNS architecture in the short term. Previous research efforts have focused on interim solutions, i.\,e., add-ons and tools that enable users who care for privacy to protect themselves against profiling and tracking efforts. The objective consists in designing and evaluating suitable privacy enhancing techniques in such a way that users do not have to rely on or trust the existing DNS infrastructure. The ``range query'' scheme by Zhao et al. \cite{Zhao:2007a} is one of those efforts.
The basic idea consists in \emph{query obfuscation}, i.\,e., sending a set of dummy queries (hence the term ``range'') with random hostnames along with the actual DNS query to the resolver.

So far the security of range query schemes has only been analyzed within a simplistic theoretical model that considers the obtainable security for \emph{singular queries}. In this paper we study the security offered by range queries for a more complex real-world application, namely \emph{web surfing}, which is one of the use cases Zhao et al. envision in \cite{Zhao:2007a}. In contrast to singular queries, downloading websites typically entails a number of inter-related DNS queries.
Our results indicate that the range query scheme offers less protection than expected in this scenario, because dependencies between consecutive queries are neglected.

The main contribution of this paper is to \textbf{demonstrate that random set range queries offer considerably less protection than expected in the web surfing use case}. We demonstrate that a curious resolver (the adversary) can launch a semantic intersection attack to disclose the actually retrieved website with high probability. We also show how the effectiveness of the attack can be reduced, and we identify a number of challenges that have to be addressed before range query schemes are suitable for practice.

The paper is structured as follows. In Sects.~2 and 3 we review existing work and fundamentals. Having described our dataset in Sect.~4, we continue with theoretical and empirical analyses in Sects.~5 and 6. We study countermeasures in Sect.~7 and discuss our results in Sect.~8. We conclude in Sect.~9.

\section{Related Work}
\label{sec:relatedWork}

The basic DNS range query scheme was introduced by Zhao et al. in \cite{Zhao:2007a}; there is also an improved version \cite{Zhao:2007b} inspired by private information retrieval \cite{Chor:1995}.
Although the authors suggest their schemes especially for web surfing applications, they fail to demonstrate their practicability using empirical results.

Castillo-Perez and Garcia-Alfaro propose a variation of the original range query scheme \cite{Zhao:2007a} using multiple DNS resolvers in parallel \cite{Castillo-Perez:2008,Castillo-Perez:2009}. They evaluate its performance for ENUM and ONS, two protocols that store data within the DNS infrastructure. Finally, Lu and Tsudik propose PPDNS \cite{Lu:2010}, a privacy-preserving resolution service that relies on CoDoNs \cite{RamasubramanianS04-codons}, a next-generation DNS system based on distributed hashtables and a peer-to-peer infrastructure, which has not been widely adopted so far.

The aforementioned publications study the security of range queries for singular queries issued independently from each other. In contrast, \cite{FederrathFHP11-dnsmixes} observes that consecutively issued queries that are dependent on each other have implications for security. They describe a timing attack that allows an adversary to determine the actually desired website and show that consecutive queries have to be serialized in order to prevent the attack. 

\section{Fundamentals}
\label{sec:fundamentals}


\subsection{Random Set DNS Range Query Scheme}
\label{sec:dnsrq}


In this paper we focus on the basic ``random set'' DNS range query scheme as introduced in \cite{Zhao:2007a}.
Zhao et al. stipulate that each client is equipped with a large database of valid domain names \textbf{(dummy database)}. Each time the client wants to issue a DNS query to a resolver, it randomly draws (without replacement) $N-1$ \emph{dummy names} from the database, and sends $N$ queries to the resolver in total. When all replies have been received from the resolver, the replies for the dummy queries are discarded and the desired reply is presented to the application that issued the query.

Zhao et al. claim that this strategy leaves the adversary with a chance of $\frac{1}{N}$ to guess the desired domain name. The value of $N$ is a security parameter, which is supposed to be chosen according to the user's privacy expectations and performance needs.



\subsection{Query Patterns}
\label{sec:patterns}

The semantic intersection attack exploits the fact that typical websites embed content from multiple servers, causing clients to issue a burst of queries for various domain names in a deterministic fashion, whenever they visit the site. For example, visiting \emph{google.com} will also trigger a DNS request for \emph{ssl.gstatic.com}, as the site includes some resources from that domain. We call the set of domain names that can be observed upon visiting a site its \emph{query pattern} $p$, i.\,e., $p(\mathrm{google.com}) = \{\mathrm{google.com},\mathrm{ssl.gstatic.com}\}$. In Sect.~\ref{sec:dataset}, we will show that many popular websites do have query patterns that can be used for this attack.


Using range queries, each individual query from a pattern $p$ is hidden in a set of $N-1$ randomly chosen queries, leading to $|p|$ sets, each containing $N$ queries, being sent to the resolver in order to retrieve all the domain names required to visit the corresponding website. We refer to $N$ as the \emph{block size} of the range query scheme and to each individual range query as a \emph{block}.

Note that the client uses standard DNS queries to deliver the range query, because it uses a conventional DNS resolver, i.\,e., a single range query with a block size of $N$ causes $N$ individual DNS queries.

\subsection{The Semantic Intersection Attack}
\label{sec:attack}

An adversary, who is in possession of a database that contains the query patterns for a set of websites he is interested in \textbf{(pattern database)}, can check whether one of these patterns can be matched to consecutive query blocks received by the client. As all the dummy names are drawn independently from each other from the dummy database, it is quite unlikely that the client will draw the pattern of a different website by chance. Therefore, the adversary can be optimistic that he will only find a single pattern in the set of consecutive range queries he receives from the client, i.\,e., the pattern of the actually desired website.

\begin{figure}[t]
\centering
\includegraphics[width=1.0\textwidth]{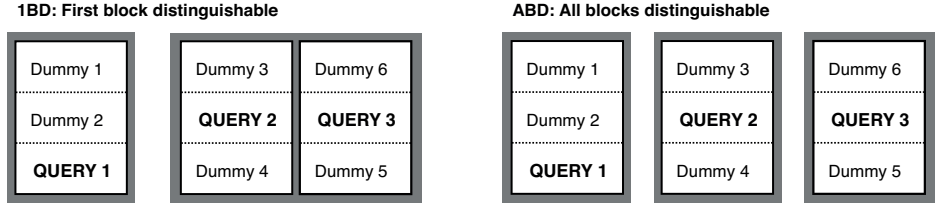}
\caption{Distinguishability of blocks for the resolver}
\label{fig:attack:1}
\end{figure}

From the viewpoint of the adversary there are two different scenarios, depending on how well the adversary can distinguish consecutive blocks (cf. Fig.~\ref{fig:attack:1}). The adversary may either be able to identify all the queries that belong to the first block, but be unable to determine which of the remaining queries belongs to which of the remaining blocks (\textbf{1BD}, 1st block distinguishable), or be able to distinguish all individual blocks, i.\,e., be able to determine for all queries to which block they belong (\textbf{ABD}, all blocks distinguishable).
The difference between the 1BD and the ABD scenario becomes evident by considering the following example. When a user visits the site \emph{\url{http://www.rapecrisis.org.uk}}, her browser will issue a query for \emph{www.rapecrisis.org.uk}. Moreover, it will issue two additional queries, for \emph{twitter.com} and \emph{www.rapecrisislondon.org}, once the HTML page has been parsed. For illustrative purposes we assume that range queries with $N=3$ are used. In the \textbf{ABD scenario} the adversary might, for instance, observe a first block of queries for (\emph{cnn.com},  \emph{www.rapecrisis.org.uk}, \emph{img.feedpress.it}), then a second  block for (\emph{github.com}, \emph{twitter.com},  \emph{s.ebay.de}), and finally a third block for  (\emph{www.rapecrisislondon.org},  \emph{ytimg.com},  \emph{conn.skype.com}). In contrast, in the \textbf{1BD scenario} the adversary might observe a first block with (\emph{cnn.com}, \emph{www.rapecrisis.org.uk}, \emph{img.feedpress.it}) and a second block with (\emph{github.com}, \emph{twitter.com}, \emph{www.rapecrisislondon.org}, \emph{s.ebay.de}, \emph{ytimg.com}, \emph{conn.skype.com}).

The first block is distinguishable in both scenarios, because the web browser has to resolve the \emph{primary domain name} in order to learn the IP address of the main web server. This IP address is received within the replies that belong to the first block of queries. After the browser has downloaded the HTML file from the main web server, it will issue queries for the \emph{secondary domain names} in order to retrieve all embedded content hosted on other web servers. 

Given a \emph{pattern database DB} that contains primary and secondary domain names of websites, the adversary proceeds as follows in order to carry out the intersection attack in the \textbf{ABD scenario}:

\begin{enumerate}
\item  From \emph{DB} the adversary selects all patterns, whose  primary domain name is contained in the first block, obtaining the set of candidates $C$.
\item The adversary selects all patterns with length $|p|$, which is the number of observed blocks, from $C$ to obtain $C_{|p|}$.
\item For each pattern $q$ in $C_{|p|}$ the adversary performs a \emph{block-wise set intersection}: $q$ is a \emph{matching pattern}, if all of its domain names are dispersed among the blocks in a plausible fashion, i.\,e., iff
  \begin{enumerate} 
    \item each block contains at least 1 element from $q$, and
    \item each element of $q$ is contained in at least 1 block, and
    \item $q$ can be completely assembled by drawing one element from each block.
  \end{enumerate}
\end{enumerate}

In the \textbf{1BD scenario} the adversary has to use a different approach, because there are only two blocks observable:

\begin{enumerate}
\item  From the pattern database the adversary selects all patterns, whose  primary domain name is contained in the first block, thus obtaining the set of candidate patterns $C$.
\item For each pattern $q$ in $C$ the adversary performs a \emph{block-wise set intersection}: $q$ is a \emph{matching pattern}, if all of its secondary domain names are contained within the second block.
\end{enumerate}

Note that due to caching, the adversary cannot reliably determine $|p|$ in the 1BD scenario. Due to variations in the lookup time of different domain names, the stub resolver on the client may already receive replies (and cache the results) for some domain names before all range queries have been submitted to the resolver. However, if the range query client happens to draw one of the cached domain names as a dummy, the stub resolver will not send another query, but answer it immediately from its cache. As a result, some queries will not reach the adversary and the effective size of consecutive blocks will vary. Therefore, the adversary cannot easily determine $|p|$ in the 1BD scenario in order to filter the set $C$. For now, we neglect the fact that caching may also affect the desired queries (cf. Sect.~\ref{sec:discussion} for a discussion of this issue).

In the remainder of the paper we focus on the \textbf{1BD scenario}, which  we deem to be more realistic than the ABD scenario. Contemporary web  browsers issue the queries for the secondary queries in parallel. Thus, when the range query client constructs range queries for each of the desired domain names, the individual queries of all the blocks will be interleaved, causing uncertainty about the composition of the individual blocks.
On the other hand, the ABD scenario is relevant for range query schemes that submit all queries contained in a block in a single message. We will consider the effect of this approach in Sect.~\ref{sec:evaluation:ex3}.


\section{Dataset}
\label{sec:dataset}

\begin{figure}[t]
\centering

\includegraphics[width=0.48\textwidth]{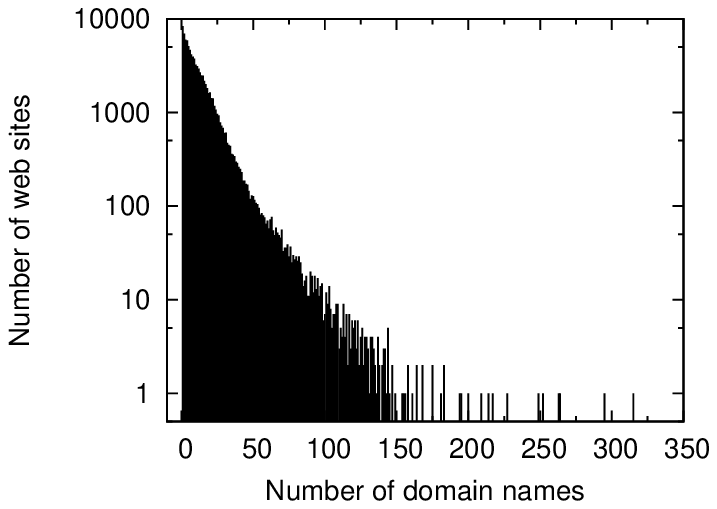}
\hfill
\includegraphics[width=0.48\textwidth]{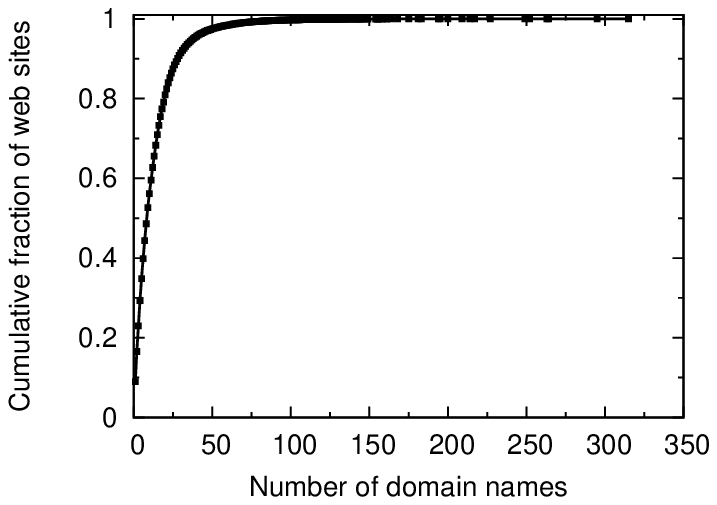}
\caption{Histogram and cumulative distribution of pattern lengths}
\label{fig:patternlengths}
\end{figure}

In order to evaluate the feasibility of the semantic intersection attack, we performed probabilistic analyses and implemented a simulator that applies the attack to the patterns of actual websites.
For this purpose we obtained the query patterns of the top $100{,}000$ websites of the ``Alexa Toplist'' (\url{http://www.alexa.com}) with the headless Webkit-based browser PhantomJS (\url{http://phantomjs.org}).\footnote{\label{fnote:github}The source code of our crawler and simulator as well as all experimental data is available at \url{https://github.com/Semantic-IA}} As PhantomJS was not able to reach and retrieve all of the websites contained in the Toplist at the time of the data collection (May 2013) the cleaned dataset contains with $|P|=92{,}880$ patterns and $|Q|=216{,}925$ unique queries. The average pattern length (\textit{mean value}) is $13.02$ with a standard deviation of $14.28$.
The distribution of pattern lengths as displayed in Fig.~\ref{fig:patternlengths} shows that, while patterns of the length 1 are frequent, patterns of higher lengths make up the majority of the dataset. The longest pattern consists of $315$ queries.

\section{Probabilistic Analysis}
\label{sec:analysis}

Before we carry out any practical evaluation using our simulator, we want to get an expectation of the likelihood of \textbf{ambiguous results}, which occur if the client happens to draw all the domain names of another website from the dummy database while the range queries needed for the desired website are assembled. If the client draws all domain names of a different pattern by chance and distributes the individual names among the blocks in a plausible fashion, the adversary will observe two patterns: the pattern of the actually desired website as well as the \textbf{random pattern}.

\subsection{Modeling the Probability of Ambiguous Results} 
\label{sec:themodel}

In the 1BD scenario an ambiguous result occurs if the primary domain name of a random pattern (the domain name of the corresponding website) is selected as a dummy in the first block, and all remaining elements of the pattern are contained in the union of the remaining blocks.\footnote{In the 1BD scenario the query distribution between the remaining blocks is irrelevant, as long as all needed queries occur at least once in the union of the blocks.}
The probability for an ambiguous result can be modeled as a series of hypergeometric distributions. A hypergeometric distribution $h(k|N;M;n)$ describes the probability of drawing $k$ elements with a specific property when drawing $n$ elements out of a group of $N$ elements, of which $M$ have the desired property:

\begin{equation}
\label{eq:analysis:1}
h(k|N;M;n) := \frac{{M \choose k}{N-M \choose n-k}}{{N \choose n}}
\end{equation}

First, we need to obtain the probability to draw the first element of a pattern of the correct length $n$ into the first block of queries. As the variables of the hypergeometric distribution overlap with those we use to describe the properties of a range query, we substitute them for their equivalents in our range query notation. $N$ is equal to $|Q|$, the number of names in the dummy database. $M$ equals to the number of patterns of the correct length, which we will write as $|P_n|$. In our case, the parameter $n$ of the hypergeometric distribution corresponds to $N-1$, as we will draw $N-1$ dummy names into the first block. By substituting these values into Eq.~\ref{eq:analysis:1}, we obtain the probability $p(n, k)$ of drawing exactly $k$ beginnings of patterns of the length $n$:

\begin{equation}
\label{eq:analysis:2}
p(n, k) := \frac{{|P_n| \choose k}{|Q|-|P_n| \choose (N-1)-k}}{{|Q| \choose N-1}}
\end{equation}

In addition to that, we need to determine the probability of drawing the remaining $k*(n-1)$ queries into the second block, which contains the remaining $(n-1)*(N-1)$ randomly drawn dummy names in the 1BD scenario. To complete our $k$ patterns, we need to draw $k*(n-1)$ specific dummy names. The probability of success is described by the function $q(n,k)$, which is given in Eq.~\ref{eq:analysis:3}.

\begin{equation}
\label{eq:analysis:3}
q(n, k) := \frac{{n-1 \choose n-1}^k {|Q|-(n-1)*k \choose (n-1)*(N-1)-(n-1)}}{{|Q| \choose (n-1)*(N-1)}} = \frac{{|Q|-(n-1)*k \choose (n-1)*(N-1)-(n-1)*k}}{{|Q| \choose (n-1)*(N-1)}}
\end{equation}

The two probabilities $p(n,k)$ and $q(n,k)$ can now be combined to receive the probability of drawing $k$ complete patterns of the correct length $n$:

\begin{equation}
\label{eq:analysis:4}
P(n, k) := p(n,k)*q(n,k)
\end{equation}

In this context, the expected value of $P(n,k)$ for different values of $n$ is of interest, as it describes the average number of patterns we expect to see. The expected value, in general, is defined as:

\begin{equation}
\label{eq:analysis:5}
E(X) := \sum\limits_{i \in I}(x_i p_i)
\end{equation}

In our case, $x_1$ is $k$, as it describes the number of patterns, and $p_i$ equals $P(n,k)$ as the probability of drawing $k$ patterns, i.\,e., the expected value is 

\begin{equation}
\label{eq:analysis:6}
E(n) := 1 + \sum\limits_{k=1}^{N-1} (P(n,k)*k)
\end{equation}

We are adding 1 to the result, as the original pattern will always be present.
Equation~\ref{eq:analysis:6} will only calculate the expected value for patterns of a specific length. However, as the adversary does not know the length of the pattern with certainty in the 1BD scenario, we have to consider patterns of any length. For that, we have to use a modified variant of Eq.~\ref{eq:analysis:3}:

\begin{equation}
\label{eq:analysis:7}
q(n, k, M) := \frac{{|Q|-(n-1)*k \choose (M-1)*(N-1)-(n-1)*k}}{{|Q| \choose (M-1)*(N-1)}}
\end{equation}

In Eq.~\ref{eq:analysis:7}, $n$ is the length of the random pattern, while $M$ is the length of the original pattern. Accordingly, we  modify Eq.~\ref{eq:analysis:4} and Eq.~\ref{eq:analysis:6}: 

\begin{equation}
\label{eq:analysis:8}
P(n, k, M) := p(n,k)*q(n,k,M)
\end{equation}

\begin{equation}
\label{eq:analysis:9}
E(M) := 1 + \sum\limits_{n=1}^M\sum\limits_{k=1}^{N-1} (P(n,k,M)*k)
\end{equation}

Finally, to determine the expected mean value of the number of detected patterns given a specific block size $N$, we calculate

\begin{equation}
\label{eq:analysis:10}
F(N) = \frac{1}{|P|}*\sum\limits_{M=1}^{L}(E(M)*|P_M|)
\end{equation}

where $L$ is the length of the longest pattern, and $|P_M|$ the number of patterns having length $M$.

\subsection{Analytical Result}
\label{sec:analyticalresult}

\begin{table}[t]
\centering
\caption{Expected avg. number of detected patterns $F(N)$ for varying block sizes $N$}
\
\begin{tabular*}{0.4\textwidth}{@{\extracolsep{\fill}}rrrr}
\toprule
$N$ & $10$ & $50$ & $100$ \\
\midrule
$F(N)$ & $1.35$ & $2.93$ & $4.83$\\
\bottomrule
\end{tabular*}
\label{tab:analysis:1}
\end{table}

The results (cf. Table~\ref{tab:analysis:1}) indicate that an adversary will, on average, detect only very few random patterns. As expected, the privacy expectation for singular queries ($\frac{1}{N}$) does not apply to the web surfing scenario.

Note that for reasons of conciseness we have provided a slightly simplified model, which disregards overlaps between patterns. Actually, the adversary must expect to find a slightly \emph{higher} number of patterns, because a domain name that is contained within multiple patterns only has to be drawn once to be detected as part of all patterns.
Nevertheless, the analysis is instructive and provides us with a baseline for the empirical evaluations that we will describe in the following.

\section{Evaluation}
\label{sec:evaluation}


In order to evaluate the effectiveness of the semantic intersection attack in a realistic scenario, we developed a simulator that enables us to efficiently test different attack strategies and various assumptions about the knowledge of the adversary. In the following we present results for the 1BD scenario.

\paragraph{Methodology}

Given a dataset the simulator will generate range queries for all the patterns from the dataset and perform the semantic intersection attack. 
We are interested in the influence of two factors on the effectiveness of the attack, namely the \emph{block size} $N$, and the \emph{size of the dummy database} $|Q|$ that contains the dummy names. If the range query scheme was to be used in practice, these two factors could be easily influenced by the user. Thus, it is worthwhile to analyze their effect on the attainable privacy. 

In the following, we will use the metric of \textbf{$k$-identifiability}, which is derived from the well-known metric $k$-anonymity \cite{Sweene02-kanonymity}: A set of consecutively observed range queries is said to be $k$-identifiable, if the adversary finds \emph{exactly} $k$ matching patterns of websites in his pattern database. For conciseness we will show the cumulative distribution of the fraction of $k$-identifiable patterns, i.\,e., the fraction of patterns that are $k$-identifiable or less than $k$-identifiable. 


\subsection{Results of Experiment 1: Variation of Block Size}
\label{sec:evaluation:ex1}

For the purpose of this analysis, we consider three different  block sizes: $N=10$, $N=50$, and $N=100$. \cite{FederrathFHP11-dnsmixes} has shown that the median latency exceeds 1200\,ms for a block size of $N=100$, rendering higher values impractical for practical use.

Based on the result of Sect. \ref{sec:analysis}, we expect to receive some, but not many ambiguous results, i.\,e., instances where the whole pattern of a different website appears in a set of consecutively observed range queries by chance. Intuitively, the larger the block size, the more random patterns will occur. Accordingly, we expect the effectiveness of the attack to degrade with increasing block sizes.

\afterpage{%
\clearpage\clearpage 

\begin{table}[t]
\centering
\caption{Results for varying block sizes $N$ given the whole dummy database}
\begin{tabular*}{0.9\textwidth}{@{\extracolsep{\fill}}rrrrrr}
\toprule
$N$ & $S$ & 1-identifiable & $\leq 5$-identifiable & median(k) & max(k) \\
\midrule
$10$ & $216{,}925$ & $62\,\%$ & $99\,\%$ & $1$ & $6$ \\
$50$ & $216{,}925$ & $8\,\%$ & $88\,\%$ & $3$ & $14$ \\
$100$ & $216{,}925$ & $1\,\%$ & $43\,\%$ & $6$ & $18$ \\
\bottomrule
\end{tabular*}
\label{tab:evaluation:ex1:1}
\end{table}

\begin{figure}[h]
\centering

\includegraphics{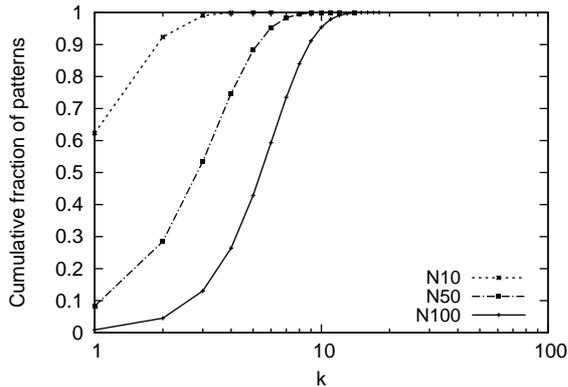}
\caption{Distribution of $k$-identifiability for varying block sizes $N$ (whole database)}
\label{fig:evaluation:ex1:1}
\end{figure}
}    

As can be seen in Table~\ref{tab:evaluation:ex1:1} and Fig.~\ref{fig:evaluation:ex1:1}, the smallest block size provides little privacy, with $62\,\%$ of patterns being 1-identifiable. Consequently, the median of the observed $k$-identifiability values is $1$. $99\,\%$ of patterns are 5-identifiable or better. No pattern is more than 6-identifiable.
For a larger block size of $N=50$, only $8\,\%$ of patterns are 1-identifiable, but the cumulative distribution quickly approaches $100\,\%$. All patterns are 14-identifiable or less, and the median of all observed $k$-identifiability values is $3$, i.\,e., for $50\,\%$ of the websites the adversary can narrow down the actually desired site to a set of 3 or less sites, which is far smaller than the baseline probability of $\frac{1}{50}$ for finding the desired domain name in the first block.
As expected, $N=100$ is most effective: $0.8\,\%$ of patterns are 1-identifiable, but still $43\,\%$ of patterns are at most 5-identifiable. 

Generally, we can observe diminishing returns when the block size is increased. While the increase from $N=10$ to 50 leads to $54\,\%$ less 1-identifiable patterns, adding another 50 queries per block only decreases the fraction by $7.2$ percentage points. The same is true for the maximum $k$-identifiability, which increases by eight and four, respectively. 
On overall, the results indicate that range queries provide far less privacy than suggested by Zhao et al. in the web surfing scenario.

\paragraph{1BD-improved}

We also considered an improved attack algorithm that guesses the length of the desired patterns based on the total number of observed queries in the second block, resulting in a range of possible pattern lengths. This allows the adversary to reject all patterns that do not fall into this range. As a result $80\,\%$ ($N=100$) and $94\,\%$ ($N=10$) of all patterns are 1-identifiable. Due to space constraints, we are unable to adequately cover the calculations to estimate the length in this paper, but we have released an implementation including the relevant documentation in the source code repository (see \Cref{fnote:github}).

\subsection{Results of Experiment 2: Variation of Dummy Database}

\afterpage{%
\clearpage\clearpage 

\begin{table}[t]
\centering
\caption{Results for varying dummy database sizes $S$ given the block size $N=50$}
\begin{tabular*}{0.9\textwidth}{@{\extracolsep{\fill}}rrrrrr}
\toprule
$N$ & $S$ & 1-identifiable & $\leq 5$-identifiable & median(k) & max(k) \\
\midrule
$50$ & $2{,}000$ & $19\,\%$ & $92\,\%$ & $3$ & $14$ \\
$50$ & $20{,}000$ & $16\,\%$ & $95\,\%$ & $3$ & $11$ \\
$50$ & $200{,}000$ & $9\,\%$ & $88\,\%$ & $3$ & $13$ \\
\bottomrule
\end{tabular*}
\label{tab:evaluation:ex2:1}
\end{table}

\begin{figure}[h]
\centering

\includegraphics{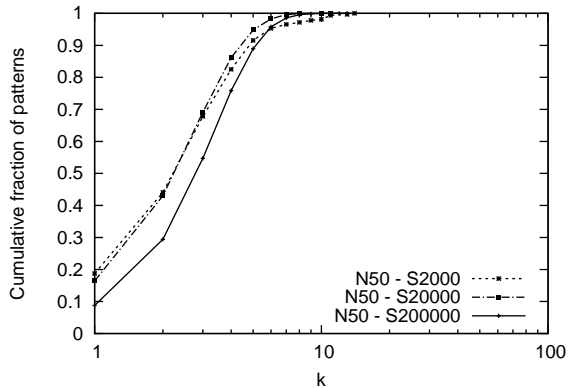}
\caption{Distribution of $k$-identifiability for varying dummy database sizes $S$; $N=50$}
\label{fig:evaluation:ex2:1}
\end{figure}
}    

Generating and maintaining a dummy database is a non-trivial task for the client, which gets harder the larger the database is supposed to be. Accordingly, the importance of the size of the dummy database is of interest.
We assume that the client's dummy database is always a subset of the pattern database of the adversary, because, in general, the adversary will have access to more resources than the client, and collecting patterns scales very well. 

We compare the effectiveness of three different database sizes ($S=2000$, $20{,}000$ and $200{,}000$). The domain names are chosen by drawing patterns from the full pattern database (without replacement) and adding all domain names of each pattern to the dummy database. This process continues until exactly $S$ unique domain names have been found. We select full patterns to increase the chance that the client randomly chooses a full pattern when drawing dummies. We used a fixed block size of $N=50$ for this experiment.

Fig.~\ref{fig:evaluation:ex2:1} shows that the differences are quite small on overall. Thus, the biggest effect of varying the database is the change in the percentage of 1-identifiable patterns: The percentage of 1-identifiable patterns drops by three percentage points when the dummy database size is increased from $S=2000$ to $S=20{,}000$, and by another 7 points on the second increase to $S=200{,}000$. 
The observed changes have a much smaller effect than the variation of the block size; however, regardless of these results, a larger database is always desirable to prevent other attacks, such as the enumeration of the client's database.





\subsection{Effect of Pattern Length on Site Identifiability}
Now that we know the effect of varying the block size, the composition of the different $k$-identifiabilities is of interest. With this information, we can determine \textbf{whether websites with longer or shorter patterns are more at risk} to be identified. Intuitively, shorter patterns should generally have lower $k$-identifiabilities, as comparatively few dummies are drawn to obfuscate them, decreasing the chance of drawing a whole pattern. Conversely, longer patterns should generally achieve higher $k$-identifiabilities, as they use a higher number of dummy domain names.
We will now test this hypothesis by analyzing the composition of the different $k$-identifiabilities, using the results of our simulation with a block size of $N=50$ and the full dummy database ($S=216{,}925$).

\begin{table}[t]
\centering
\caption{Number of patterns $n_k$, mean length $\overline{|p|}$ and standard deviation $\mathrm{SD}$ aggregated by resulting $k$-identifiability ($N=50$, $S=216{,}925$)}
\label{tab:evaluation:reasons:1}
\begin{tabular*}{1\textwidth}{@{\extracolsep{\fill}}rrrrrrrrrrr}
\toprule
$k$ & $1$ & $2$ & $3$ & $4$ & $5$ & $6$ & $7$ & $8$ & $9$ & $\geq10$ \\
\midrule
$n_k$ & $7{,}693$ & $18{,}790$ & $23{,}184$ & $19{,}784$ & $12{,}497$ & $6{,}532$ & $2{,}875$ & $1{,}077$ & $336$ & $121$  \\
$\overline{|p|}$ & $10.59$ & $11.43$ & $12.52$ & $13.54$ & $14.43$ & $15.45$ & $16.22$ & $17.65$ & $17.09$ & $19.47$ \\
$\mathrm{SD}$ & $12.16$ & $13.24$ & $13.65$ & $14.55$ & $15.02$ & $16.14$ & $16.65$ & $17.71$ & $15.35$ & $19.68$ \\
\bottomrule
\end{tabular*}

\end{table}

As can be seen in Table~\ref{tab:evaluation:reasons:1}, the mean pattern length rises almost linearly with increasing $k$-identifiability, which supports our hypothesis.
The standard deviation exhibits a similar behavior, albeit with a slightly lower and less uniform growth rate. We could reproduce this result for other block and database sizes. The correlation is more distinct for larger block sizes. Smaller block sizes do not show this behavior as clearly, as the range of $k$-identifiabilities is too small to show any distinct trend.

\subsection{Results of Experiment 3: ABD Scenario}
\label{sec:evaluation:ex3}
So far, we concentrated on the 1BD scenario (cf. Sect.~\ref{sec:attack}). We will now consider the ABD scenario by repeating the experiment from Sect.~\ref{sec:evaluation:ex1}, simulating an adversary that can distinguish individual blocks:
In the ABD scenario the adversary is able to 1-identify between $87\,\%$ ($N=100$) and $97\,\%$ ($N=10$) of all domain names, vastly improving on the results of 1BD ($1\,\%$ and $62\,\%$, respectively).

The increased accuracy is due to two effects: Firstly, in the ABD scenario the adversary can derive $|p|$, the length of the obfuscated pattern, and filter the set of candidate patterns accordingly (cf. Sect.~\ref{sec:attack}). Secondly, the probability that another matching pattern is drawn from the dummy database by chance is much smaller when it has to meet the three ABD conditions.

The contribution of these two effects to the overall effectiveness obtained for ABD can be analyzed by reviewing the results obtained for the baseline (1BD) in comparison to 1BD-improved (cf. Sect.\ref{sec:evaluation:ex1}) and ABD: The results for 1BD-improved, which filters candidate patterns using a vague estimation of $|p|$,  already show a significant increase: For $N=50$ the fraction of 1-identifiable sites is $83\,\%$ for 1BD-improved, while it is only $8\,\%$ for 1BD. On the other hand, the fraction of 1-identifiable websites obtained for ABD, where matching patterns have to meet the additional conditions and the exact value of $|p|$ is known, rises only by another 6 percentage points (reaching $89\,\%$) compared to 1BD-improved. 

While this sort of analysis can not conclusively prove that the effect of filtering by length is larger than the effect of filtering via the ABD conditions, we note that the additional benefit of these conditions is comparatively small when the adversary can estimate the length of the obfuscated pattern.

This result indicates that range query schemes that are supposed to provide privacy in a web surfing scenario have to be devised and implemented in a way that the adversary cannot infer the length of the obfuscated query pattern. 

\section{Countermeasures}
\label{sec:countermeasures}

Having shown the weaknesses of the range query scheme against a pattern-based attack strategy, we will now discuss possible countermeasures. First, we will discuss and evaluate a pattern-based dummy selection strategy. Afterwards, we will consider other strategies that could be used to hinder the adversary.

\subsection{Pattern-Based Dummy Selection Strategy}
\label{sec:countermeasures:improved-dummy}

In the original dummy selection strategy, the client sampled the dummies independently and randomly from his dummy database. In contrast, the client will now draw \emph{whole patterns} from his database. When querying the resolver for a desired pattern, the client will draw $N-1$ random patterns of the same length and use them as dummies. If not enough patterns of the correct length are available, the client will combine two shorter patterns to obtain a concatenated pattern with the correct length. Intuitively, this approach ensures that the adversary will always detect $N$ patterns. The results of our evaluation, shown in Table~\ref{tab:countermeasures:improved-dummy:1}, confirm this conjecture. All patterns are exactly $N$-identifiable.

\begin{table}[t]
\centering
\caption{Statistics for varying block sizes $N$ using the pattern-based dummy construction strategy}
\begin{tabular*}{0.9\textwidth}{@{\extracolsep{\fill}}rrrrrr}
\toprule
$N$ & $S$ & 1-identifiable & $\leq 5$-identifiable & median(k) & max(k) \\
\midrule
$10$ & $216{,}925$ & $0\,\%$ & $0\,\%$ & $10$ & $10$ \\
$50$ & $216{,}925$ & $0\,\%$ & $0\,\%$ & $50$ & $50$ \\
$100$ & $216{,}925$ & $0\,\%$ & $0\,\%$ & $100$ & $100$ \\
\bottomrule
\end{tabular*}
\label{tab:countermeasures:improved-dummy:1}
\end{table}

However, in real-world usage scenarios, the length of the pattern the client is about to query cannot be known in advance. As the dummies for the first element of the pattern have to be chosen before the query can be sent, the client has no way to be sure of the pattern length of the desired website, as these values may change over time when a website changes. This leads to uncertainty about the correct length of the dummy patterns. A wrong choice of pattern length may be used by the adversary to identify the original pattern.
Future research could study more sophisticated dummy selection strategies, drawing from experience gained in the field of obfuscated web search \cite{BalsaTD12}.

\subsection{Other Countermeasures}

As described in the previous section the pattern-based dummy selection strategy is subject to practical limitations. We will briefly cover other countermeasures that may be used to improve the privacy of clients. This list is not exhaustive.

The first option is to use a variable value for $N$ that changes on each block. This will raise the difficulty of determining the length of the original pattern, as long as the adversary cannot distinguish individual blocks. This change would render 1BD-improved useless, as it depends on a fixed number of chosen dummies per block (although similar optimizations could be found that would still improve on the performance of the trivial algorithm). However, this would not impact the performance of the ABD algorithm, as it does not rely on uniform block sizes.

Another improvement that may make the pattern-based strategy more feasible would be to round up the length of the target pattern to the next multiple of a number $x > 1$. The additional queries (``padding'') could be chosen randomly, or by choosing patterns of the correct length.

Finally, other privacy-enhancing techniques, such as mixes and onion routing \cite{chaum81-mix,dingledine04tor}, can be employed to counter monitoring and tracking efforts. However, these general-purpose solutions are not specifically designed for privacy-preserving DNS resolution and may introduce significant delays into the resolution process.

\section{Discussion}
\label{sec:discussion}



We designed our experimental setup to stick as closely as possible to reality. However, for reasons of conciseness and clarity we have neglected some effects. In the following we will discuss whether they affect the validity of our conclusions.

Firstly, the results are implicitly biased due to a closed-world assumption, i.\,e., our results have been obtained on a dataset of limited size. However, as the Toplist of Alexa contains a large variety of websites we are confident that the results are valid for a large fraction of sites in general. Moreover, we have only evaluated the effectiveness of the attack for the \emph{home pages}; the evaluation of the attack on individual sub-pages is left for future work.

Secondly, while we considered the effects of caching of dummy queries in the 1BD scenario, we disregarded caching of the desired queries: The client may still have (parts of) a pattern in his local cache, resulting in incomplete patterns being sent to the resolver. However, the adversary may adapt to caching by remembering the TTL of all responses he sent to a client and matching the patterns against the union of the received domain names and the cached entries. 

Moreover, an adversary who wants to determine all websites a user visits needs the patterns of all websites on the Internet. Such a database would be non-trivial to generate and maintain. However, a \emph{reactive} adversary may visit any domain name he receives a query for and store the pattern for that domain name in its pattern database, making a slightly delayed identification possible.

Finally, we disregarded changing patterns as well as DNS prefetching techniques, which cause longer and more volatile patterns. However, a determined adversary will have no problems in addressing these issues.

\section{Conclusion}
\label{sec:conclusion}


We demonstrated that random set range queries offer considerably less protection than expected in the use case of web surfing. Our attack exploits characteristic query patterns, which lead to greatly reduced query privacy compared to the estimations made by Zhao et al. in their original work. Moreover, we proposed and evaluated an improved range query scheme using query patterns to disguise the original pattern.
We encourage researchers to consider the effects of semantic interdependencies between queries when designing new schemes for query privacy, as the rising pervasiveness of social networking buttons, advertising and analytics makes singular queries less and less common for web surfing.

\medskip

 
\bibliography{dnsprivacy}

\begin{thebibliography}{10}
\providecommand{\url}[1]{\texttt{#1}}
\providecommand{\urlprefix}{URL }

\bibitem{rfc4033}
Arends, R., Austein, R., Larson, M., Massey, D., Rose, S.: {DNS Security
  Introduction and Requirements}. {RFC} 4033 (Mar 2005)

\bibitem{BalsaTD12}
Balsa, E., Troncoso, C., D\'{\i}az, C.: {OB-PWS: Obfuscation-Based Private Web
  Search}. In: IEEE Symposium on Security and Privacy. pp. 491--505. IEEE
  Computer Society (2012)

\bibitem{Castillo-Perez:2008}
Castillo-Perez, S., Garc\'{\i}a-Alfaro, J.: {Anonymous Resolution of DNS
  Queries}. In: On the Move to Meaningful Internet Systems (OTM 2008).
  Proceedings, Part II. pp. 987--1000. Springer, LNCS 5332 (2008)

\bibitem{Castillo-Perez:2009}
Castillo-Perez, S., Garc\'{\i}a-Alfaro, J.: {Evaluation of Two
  Privacy--Preserving Protocols for the DNS}. In: Sixth International
  Conference on Information Technology: New Generations (ITNG 2009).
  Proceedings. pp. 411--416. IEEE, Washington, DC, USA (2009)

\bibitem{chaum81-mix}
Chaum, D.: {Untraceable electronic mail, return addresses, and digital
  pseudonyms}. {Communications of the ACM}  24(2) (1981)

\bibitem{Chor:1995}
Chor, B., Goldreich, O., Kushilevitz, E., Sudan, M.: {Private Information
  Retrieval}. In: 36th Annual Symposium on Foundations of Computer Science
  (FOCS 1995). Proceedings. pp. 41--50. IEEE Computer Society (1995)

\bibitem{Conrad12-dnssecurity}
Conrad, D.: {Towards Improving DNS Security, Stability, and Resiliency} (2012),
  \url{http://www.internetsociety.org/sites/default/files/bp-dnsresiliency-201201-en_0.pdf}

\bibitem{dingledine04tor}
Dingledine, R., Mathewson, N., Syverson, P.: {Tor: The Second-Generation Onion
  Router}. In: 13th USENIX Security Symposium. pp. 303--320 (2004)

\bibitem{rfc2535}
Eastlake, D.: {Domain Name System Security Extensions}. RFC 2535 (Mar 1999)

\bibitem{FederrathFHP11-dnsmixes}
Federrath, H., Fuchs, K.P., Herrmann, D., Piosecny, C.: {Privacy-Preserving
  DNS: Analysis of Broadcast, Range Queries and Mix-Based Protection Methods}.
  In: European Symposium on Research in Computer Security (ESORICS 2011).
  Proceedings. pp. 665--683. Springer, LNCS 6879 (2011)

\bibitem{HBF:2013}
Herrmann, D., Banse, C., Federrath, H.: {Behavior-based Tracking: Exploiting
  Characteristic Patterns in DNS Traffic}. Computers \& Security  39A,  17--33
  (Nov 2013)

\bibitem{Lu:2010}
Lu, Y., Tsudik, G.: {Towards Plugging Privacy Leaks in the Domain Name System}.
  In: Tenth International Conference on Peer-to-Peer Computing (P2P 2010).
  Proceedings. pp. 1--10. IEEE (2010)

\bibitem{RamasubramanianS04-codons}
Ramasubramanian, V., Sirer, E.G.: {The Design and Implementation of a Next
  Generation Name Service for the Internet}. In: Yavatkar, R., Zegura, E.W.,
  Rexford, J. (eds.) SIGCOMM 2004 Conference. Proceedings. pp. 331--342. ACM
  (2004)

\bibitem{Sweene02-kanonymity}
Sweeney, L.: {k-Anonymity: A Model for Protecting Privacy}. International
  Journal of Uncertainty, Fuzziness and Knowledge-Based Systems  10(5),
  557--570 (2002)

\bibitem{Zhao:2007a}
Zhao, F., Hori, Y., Sakurai, K.: {Analysis of Privacy Disclosure in DNS Query}.
  In: International Conference on Multimedia and Ubiquitous Engineering (MUE
  2007). Proceedings. pp. 952--957. IEEE (2007)

\bibitem{Zhao:2007b}
Zhao, F., Hori, Y., Sakurai, K.: {Two--Servers PIR Based DNS Query Scheme with
  Privacy--Preserving}. In: International Conference on Intelligent Pervasive
  Computing (IPC 2007). Proceedings. pp. 299--302. IEEE (2007)

\bibitem{Zhao:2010}
Zhao, F., Hori, Y., Sakurai, K.: {Analysis of Existing Privacy--Preserving
  Protocols in Domain Name System}. IEICE Transactions  93-D(5),  1031--1043
  (2010)

\end{thebibliography}
\bibliographystyle{splncs03}
\end{document}